\documentclass[10pt,conference,USletter]{IEEEtran}
\IEEEoverridecommandlockouts

\usepackage{graphicx}
\usepackage{url} 

\usepackage{amsmath}  
\usepackage{amsfonts} 
\usepackage{amsthm}   
\usepackage{amsbsy}   
\usepackage{amssymb}  
\usepackage{csquotes} 

\usepackage{cite}
\usepackage{algorithm}
\usepackage{algpseudocode}
\usepackage{pgfplots}
\algnewcommand{\FForAll}[1]{\State\algorithmicforall\ #1\ \algorithmicdo}
\algnewcommand{\EEndFor}{\unskip\ \algorithmicend\ \algorithmicfor}

\usepackage{tikz}

\usetikzlibrary{plotmarks}
\pgfplotsset{compat=newest}
\pgfplotsset{plot coordinates/math parser=false}
\usetikzlibrary{matrix,calc,backgrounds}

\makeatletter
\def\squarecorner#1{
    \pgf@x=\the\wd\pgfnodeparttextbox%
    \pgfmathsetlength\pgf@xc{\pgfkeysvalueof{/pgf/inner xsep}}%
    \advance\pgf@x by 2\pgf@xc%
    \pgfmathsetlength\pgf@xb{\pgfkeysvalueof{/pgf/minimum width}}%
    \ifdim\pgf@x<\pgf@xb%
    \pgf@x=\pgf@xb%
    \fi%
    \pgf@y=\ht\pgfnodeparttextbox%
    \advance\pgf@y by\dp\pgfnodeparttextbox%
    \pgfmathsetlength\pgf@yc{\pgfkeysvalueof{/pgf/inner ysep}}%
    \advance\pgf@y by 2\pgf@yc%
    \pgfmathsetlength\pgf@yb{\pgfkeysvalueof{/pgf/minimum height}}%
    \ifdim\pgf@y<\pgf@yb%
    \pgf@y=\pgf@yb%
    \fi%
    \ifdim\pgf@x<\pgf@y%
    \pgf@x=\pgf@y%
    \else
    \pgf@y=\pgf@x%
    \fi
    \pgf@x=#1.5\pgf@x%
    \advance\pgf@x by.5\wd\pgfnodeparttextbox%
    \pgfmathsetlength\pgf@xa{\pgfkeysvalueof{/pgf/outer xsep}}%
    \advance\pgf@x by#1\pgf@xa%
    \pgf@y=#1.5\pgf@y%
    \advance\pgf@y by-.5\dp\pgfnodeparttextbox%
    \advance\pgf@y by.5\ht\pgfnodeparttextbox%
    \pgfmathsetlength\pgf@ya{\pgfkeysvalueof{/pgf/outer ysep}}%
    \advance\pgf@y by#1\pgf@ya%
}
\makeatother

\pgfdeclareshape{square}{
    \savedanchor\northeast{\squarecorner{}}
    \savedanchor\southwest{\squarecorner{-}}
    
    \inheritanchor[from=rectangle]{base east}
    \inheritanchor[from=rectangle]{base west}
    \inheritanchor[from=rectangle]{base}
    \inheritanchor[from=rectangle]{center}
    \inheritanchor[from=rectangle]{east}
    \inheritanchor[from=rectangle]{mid east}
    \inheritanchor[from=rectangle]{mid west}
    \inheritanchor[from=rectangle]{mid}
    \inheritanchor[from=rectangle]{north east}
    \inheritanchor[from=rectangle]{north west}
    \inheritanchor[from=rectangle]{north}
    \inheritanchor[from=rectangle]{south east}
    \inheritanchor[from=rectangle]{south west}
    \inheritanchor[from=rectangle]{south}
    \inheritanchor[from=rectangle]{west}
    \inheritanchorborder[from=rectangle]
    \inheritbackgroundpath[from=rectangle]
}

\usepackage[T1]{fontenc}
\usepackage[utf8]{inputenc}
\usepackage{bm}
\usepackage{balance}
\usepackage{comment}
\usepackage{lipsum}

\usepackage[capitalize]{cleveref}
  \crefname{equation}{}{}
  \crefname{thm}{Theorem}{Theorems}

\newcommand{\Reals}{\mathbb{R}}
\renewcommand{\vec}[1]{\bm {#1}}
\newcommand{\Neighbours}[2][]{\mathcal{N}_{#1}\left( #2 \right)}
\DeclareMathOperator{\supp}{supp}
\DeclareMathOperator{\IPA}{IPA}

\newtheorem{thm}{Theorem}
\newtheorem{lemma}{Lemma}
\newtheorem{proposition}{Proposition}
\newtheorem{define}{Definition}

\begin{document}

        \title{On Failing Sets of the Interval-Passing Algorithm for Compressed Sensing}
    
\author{\IEEEauthorblockN{Yauhen Yakimenka\IEEEauthorrefmark{2} and Eirik Rosnes\IEEEauthorrefmark{3}}\\
\vspace{-4mm}\IEEEauthorblockA{\IEEEauthorrefmark{2}Institute of Computer Science, University of Tartu, Tartu 50409, Estonia\\ Email: yauhen@ut.ee}
\IEEEauthorblockA{\IEEEauthorrefmark{3}Department of Informatics,
University of Bergen, N-5020 Bergen, Norway, and the Simula Research Lab\\ Email:
eirik@ii.uib.no}
\thanks{The work of Y.\ Yakimenka and E.\ Rosnes was partially funded by the Norwegian-Estonian Research Cooperation Programme (grant EMP133). The work of E.\ Rosnes was partially funded by the Research Council of Norway (grant 240985/F20) and by Simula@UiB.
    
    The calculations were carried out in part in the High Performance Computing Centre of University of Tartu.
    }}%

    	\maketitle

\begin{abstract}

In this work, we analyze the failing sets of the interval-passing algorithm (IPA) for compressed sensing. The IPA is an efficient iterative algorithm for reconstructing a $k$-sparse nonnegative $n$-dimensional real signal $\vec{x}$ from a small number of linear measurements $\vec{y}$. In particular, we show that the IPA fails to recover $\vec{x}$ from $\vec{y}$ if and only if it fails to recover a corresponding binary vector of the same support, and also that only positions of nonzero values in the measurement matrix are of importance for success of recovery. Based on this observation, we introduce \emph{termatiko sets} and show that the IPA fails to fully recover $\vec x$ if and only if the support of $\vec x$ contains a nonempty termatiko set, thus giving a complete (graph-theoretic) description of the failing sets of the IPA. Finally, we present an extensive numerical study showing that in many cases there exist termatiko sets of size strictly smaller than the stopping distance of the binary measurement matrix; even as low as half the stopping distance in some cases.
\end{abstract}

\section{Introduction}\label{sec:intro}

\IEEEPARstart{T}{he} reconstruction of a (mathematical) object from a partial set of observations in an efficient and reliable manner is of fundamental importance. Compressed sensing, motivated by the ground-breaking work of Cand\`{e}s and Tao \cite{can06,can05}, and independently by Donoho \cite{don06}, is a research area in which the object to be reconstructed is a $k$-sparse signal vector (there are at most $k$ nonzero entries in the vector) over the real numbers. The partial information provided is a linear transformation of the signal vector, the \emph{measurement vector}, and the objective is to reconstruct the object from a small number of measurements. Compressed sensing provides a mathematical framework which shows that, under some conditions, signals can be recovered from far less measurements than with conventional signal acquisition methods. The main idea in compressed sensing is to exploit that most interesting signals have an inherent structure or contain redundancy.

Iterative reconstruction algorithms for compressed sensing have received considerable interest recently. See, for instance, %
\cite{zha12,RDVD12,cha10,pha09,Don13} and references therein. The interval-passing algorithm (IPA) for reconstruction of nonnegative sparse signals was introduced by Chandar \emph{et al.} in \cite{cha10} for binary measurement matrices. The algorithm was further generalized to nonnegative real measurement matrices in \cite{RDVD12}.

In this work, we show that the IPA fails for a nonnegative signal $\vec{x} =(x_1,\dotsc,x_n) \in \Reals_{\geq 0}^n$, $\Reals_{\geq 0}$ is the set of nonnegative real numbers, if and only if it fails for a corresponding binary vector $\vec{z}$ of the same support,  and also that only positions of nonzero values in the measurement matrix are of importance for success of recovery. Thus, failing sets as subsets of $[n] \triangleq \{1,\dotsc,n\}$ can be defined. It has previously been shown that traditional stopping sets for belief propagation decoding of low-density parity-check (LDPC) codes are failing sets of the IPA, in the sense that if the support of a signal $\vec x \in \Reals_{\geq 0}^n$ contains a nonempty stopping set, then the IPA fails to fully recover $\vec x$ \cite[Thm.~1]{RDVD12}. In this work, we extend the results in \cite{RDVD12} and define \emph{termatiko sets} (which contain stopping sets as a special case) and show that the IPA fails to fully recover a signal $\vec x \in \Reals_{\geq 0}^n$  if and only if the support of $\vec x$ contains a nonempty termatiko set, thus giving a complete (graph-theoretic) description of the failing sets of the IPA. Finally, we present an extensive numerical study which includes both specific binary parity-check matrices of LDPC codes and parity-check matrices from LDPC code ensembles as measurement matrices. The numerical results show that in many cases there exist termatiko sets of size strictly smaller than the stopping distance of the measurement matrix; even as low as half the stopping distance in some cases,  where the stopping distance $s_{\rm min}$ of a measurement matrix is the minimum size of a nonempty stopping set of the matrix when it is regarded as a parity-check matrix of a binary LDPC code.

We remark that the performance of the IPA and its comparison with other algorithms for efficient reconstruction of sparse signals have been investigated in \cite{RDVD12} (see Figs.~4 and 8), and we refer the interested reader to that work for such results. %

\section{Notation and Background}
In this section, we introduce the problem formulation, revise notation from \cite{RDVD12}, and describe the IPA in detail. 

\subsection{Compressed Sensing}
	Let $\vec{x} \in \Reals^n$, where $\Reals$ is the field of real numbers, be an $n$-dimensional 
	$k$-sparse signal (i.e., it has at most $k$ nonzero entries), and let $A = (a_{ji})$  be an  $m \times n$ real measurement matrix. We consider the recovery of $\vec{x}$ from 
	measurements $\vec{y} = A \vec{x} \in \Reals^m$, where $m < n$ and $k < n$. 

	The reconstruction problem of compressed sensing is to find the sparsest $\vec{x}$ (or the one that minimizes the $\ell_0$-norm) under the constraint $\vec{y} = A \vec{x}$, which in general is an NP-hard problem. Basis pursuit is an algorithm which reconstructs $\vec{x}$ by minimizing its $\ell_1$-norm under the constraint $\vec{y} = A \vec{x}$ \cite{can05}. This is a linear program, and thus it can be solved in polynomial time. The algorithm has a remarkable performance, but its complexity is high, making it impractical for many applications that require fast reconstruction. A fast reconstruction algorithm for nonnegative real signals and measurement matrices is the IPA which is described below in \cref{sec:ipa}.

\subsection{Tanner Graph Representation}
    We associate with matrix $A$ the bipartite Tanner graph $G = (V \cup C, 
    E)$, where $V = \{ v_1, v_2, \dotsc, v_n \}$ is a set of \emph{variable 
    nodes}, $C = \{ c_1, c_2, \dotsc, c_m \}$ is a set of \emph{measurement 
    nodes}, and $E$ is a set of edges from $C$ to $V$. We will often equate $V$ with $[n]$ and $C$ with $[m]$. There is an edge in $E$ between $c \in C$ and $v \in V$ if and only if 
    $a_{cv} \neq 0$. We also denote the sets of neighbors for each node $v \in V$ and $c \in C$ as
    \begin{align*}
        &\Neighbours{v} = \{ c \in C \mid (c, v) \in E \} \\
        \text{and}\quad&\Neighbours{c} = \{ v \in V \mid (c, v) \in E \} \,,
    \end{align*}
respectively.    Furthermore, if $T \subset V$ or $T \subset C$ and $w \in V \cup C$, then define
    \begin{displaymath}
        \Neighbours{T}    = \bigcup_{t \in T} \Neighbours{t} \text{ and }
        \Neighbours[T]{w} = \Neighbours{w} \cap T \,.
    \end{displaymath}
A stopping set \cite{di02} of the Tanner graph $G$ is defined as a subset $S$ of $V$ such that all its neighboring measurement nodes are connected at least twice to $S$. %

\subsection{Interval-Passing Algorithm} \label{sec:ipa}
The IPA is an iterative algorithm to reconstruct a nonnegative real signal $\vec x \in \Reals_{\geq 0}^n$ from a set of linear measurements $\vec y = A \vec x$, introduced by Chandar \emph{et al.} in \cite{cha10} for binary measurement matrices.  The algorithm was extended to nonnegative real  measurement matrices in  \cite{RDVD12}, and this is the case that we will consider. The IPA iteratively sends messages between variable and measurement nodes. Each message contains two real numbers, a \emph{lower bound} and an \emph{upper bound} on the value of the variable node to which it is affiliated. Let $\mu^{(\ell)}_{v \to c}$ (resp.\ $\mu^{(\ell)}_{c \to v}$) denote the lower bound of the message from variable node $v$ (resp.\ measurement node $c$) to measurement node $c$ (resp.\ variable node $v$) at iteration $\ell$.  The corresponding upper bound of the message is denoted by  $M^{(\ell)}_{v \to c}$ (resp.\ $M^{(\ell)}_{c \to v}$). It is a distinct property of the algorithm that at any iteration $\ell$, $\mu^{(\ell)}_{v \to c} \leq x_v \leq M^{(\ell)}_{v \to c}$ and $\mu^{(\ell)}_{c \to v} \leq x_v \leq M^{(\ell)}_{c \to v}$, for all $v \in V$ and $c \in \Neighbours{v}$.

The detailed steps of the IPA are shown in \cref{alg:ipa} below. From \cref{alg:ipa:muM0,alg:ipa:muvc,alg:ipa:Mvc} one can see that both $\mu^{(\ell)}_{v \to c}$ and $M^{(\ell)}_{v \to c}$ are independent of $c \in \Neighbours v$. Thus, we will occasionally denote $\mu^{(\ell)}_{v \to c}$  by $\mu^{(\ell)}_{v \to \cdot}$ and $M^{(\ell)}_{v \to c}$ by $M^{(\ell)}_{v \to \cdot}$.

    \begin{algorithm}
        \caption{Interval-Passing Algorithm (cf.\ \cite[Alg.~1]{RDVD12})}
        \label{alg:ipa}
        
        \begin{algorithmic}[1]
            \Function {IPA}{$\vec{y}$, $A$}
                \Statex Initialization
                \ForAll {$v \in V$, $c \in \Neighbours{v}$} 
                    \State $\begin{aligned}
                             \mu^{(0)}_{v \to c} \leftarrow 0 \text{ and }
                             M^{(0)}_{v \to c} \leftarrow \min_{c' \in \Neighbours{v}} \left( y_{c'} / a_{c'v} \right) 
                           \end{aligned}$ \label{alg:ipa:muM0}
                \EndFor
                
                \Statex Iterations
                \State $\ell \leftarrow 0$
                \Repeat
                    \State $\ell \leftarrow \ell+1$
                    
                    \ForAll {$c \in C$, $v \in \Neighbours{c}$}
                    \State $\begin{aligned}
                    \mu^{(\ell)}_{c \to v} \leftarrow \frac{1}{a_{cv}}\left(y_c - \sum_{v' \in 
                            \Neighbours{c}, v' \neq v} a_{cv'} M^{(\ell-1)}_{v' \to 
                        c}\right)\end{aligned}$ \label{alg:ipa:mucv}
                    \If {$\mu^{(\ell)}_{c \to v} < 0$}
                    \State $\mu^{(\ell)}_{c \to v} \leftarrow 0$
                    \EndIf
                    \State $\begin{aligned}
                    M^{(\ell)}_{c \to v} \leftarrow \frac{1}{a_{cv}}\left( y_c - \sum_{v' \in \Neighbours{c}, v' \neq v} a_{cv'} \mu^{(\ell-1)}_{v' \to c} \right)
                           \end{aligned}$ \label{alg:ipa:Mcv}
                    \EndFor
                    
                    \ForAll {$v \in V$, $c \in \Neighbours{v}$}
                        \State $\begin{aligned}
                            \mu^{(\ell)}_{v \to c} \leftarrow \max_{c' \in \Neighbours{v}} 
                             \mu^{(\ell)}_{c' \to v}  
                               \end{aligned}$ \label{alg:ipa:muvc}
                        \State $\begin{aligned}
                            M^{(\ell)}_{v \to c} \leftarrow \min_{c' \in \Neighbours{v}} 
                             M^{(\ell)}_{c' \to v} 
                               \end{aligned}$ \label{alg:ipa:Mvc}
                    \EndFor

                \Until{$\mu^{(\ell)}_{v \to \cdot} = \mu^{(\ell-1)}_{v \to \cdot}$ and $M^{(\ell)}_{v \to \cdot} = M^{(\ell-1)}_{v \to \cdot}$, \: $\forall v \in V$}
                
                \Statex Result
                \FForAll {$v \in V$}
                    $\begin{aligned}
                        \hat{x}_v \leftarrow \mu^{(\ell)}_{v \to \cdot}
                    \end{aligned}$ 
                \EEndFor \label{alg:hatxv}
                \State \textbf{return} $\hat{\vec x}$
                
            \EndFunction
        \end{algorithmic}
    \end{algorithm}

\section{Failing Sets of the Interval-Passing Algorithm} \label{sec:failing_sets}
In this section, we present several results related to the failure of the IPA. In particular, in \cref{sec:thm:supp-converge}, we show that the IPA fails to recover $\vec{x}$ from $\vec{y}$ if and only if it fails to recover a corresponding binary vector of the same support, and also that only positions of nonzero values in the matrix $A$ are of importance for success of recovery (see \cref{thm:0-1-x} below). Based on \cref{thm:0-1-x}, we introduce the concept of termatiko sets in \cref{sec:termatiko_sets} and give a complete (graph-theoretic) description of the failing sets of the IPA in \cref{sec:criterion}.

\subsection{Signal Support Recovery} \label{sec:thm:supp-converge}
    Consider the two related problems IPA($\vec y$, $A$) and IPA($\vec s$, $B$), where $\vec s = B \vec z$ and $\vec z \in \{0,1\}^n$ has support $\supp(\vec z) = \supp(\vec x)$, i.e., $\vec x$ and $\vec z$ have the same support. The support of a real vector $\vec x \in \Reals^n$ is defined as the set of nonzero coordinates of $\vec x$. The binary matrix $B$ contains ones exactly in the positions where $A$ has nonzero values. We will show below (see \cref{thm:0-1-x}) that these two problems behave identically, namely they 
    recover exactly the same positions of $\vec x$ and $\vec z$. However, note that this is true if the identical algorithm (\cref{alg:ipa}) is applied to both problems, i.e., the binary nature of $\vec z$ is not exploited.

    \begin{lemma} \label{thm:0-1-x}
        Let $A = (a_{ji}) \in \Reals_{\geq 0}^{m \times n}$, $\vec x \in \Reals_{\geq 0}^n$, $B = (b_{ji}) \in \{0,1\}^{m \times n}$, and $\vec z \in \{0,1\}^n$, where $\supp(\vec z) = \supp(\vec x)$ and
        \[
       b_{ji} = \begin{cases}
       0 &\text{if $a_{ji} = 0$} \,,\\
       1 &\text{otherwise} \,.
       \end{cases}
       \]
       Further, denote $\vec y = A \vec x$, $\vec s = B \vec z$, $\hat{\vec x} = 
        \mathrm{IPA}(\vec y, A)$, and $\hat{\vec z} = \mathrm{IPA}(\vec s, B)$.  
        Then, for all $v \in V$, %
        \[
            \hat{x}_v = x_v \quad \text{if and only if} \quad \hat{z}_v = z_v 
            \,.
        \]
    \end{lemma}

    \begin{IEEEproof}
        Define subsets of $V$ in which either the lower or the upper bound of a variable-to-measurement message, at a given iteration $\ell$, is equal to $x_v$ or $z_v$ as follows: %
        \begin{align*}
            \gamma_x^{(\ell)} &= \left\{ v \in V \mid \mu^{(\ell)}_{v \to \cdot} = x_v 
            \right\},	
            \Gamma_x^{(\ell)} = \left\{ v \in V \mid M^{(\ell)}_{v \to \cdot} = x_v  
            \right\},	\\
            \gamma_z^{(\ell)} &= \left\{ v \in V \mid \lambda^{(\ell)}_{v \to \cdot} = 
            z_v \right\},	
            \Gamma_z^{(\ell)} = \left\{ v \in V \mid \Lambda^{(\ell)}_{v \to \cdot} = 
            z_v \right\},
        \end{align*}
        where $\lambda^{(\ell)}_{v \to \cdot}$ and $\Lambda^{(\ell)}_{v \to \cdot}$ denote, respectively, the lower and the upper bound of the variable-to-measurement message from variable node $v$ to any measurement node $c \in \Neighbours{v}$ 
at iteration $\ell$ for IPA($\vec s$, $B$) (analogously to $\mu^{(\ell)}_{v \to \cdot}$ and $M^{(\ell)}_{v \to \cdot}$ for IPA($\vec y$, $A$)).
        
        To prove the lemma, it is enough to show that at each iteration $\ell$, $\gamma_x^{(\ell)} = \gamma_z^{(\ell)}$ and $\Gamma_x^{(\ell)} = \Gamma_z^{(\ell)}$. We demonstrate this by induction on $\ell$. 
        
        {\flushleft\emph{Base Case}.}
        \begin{align*}
            \gamma_x^{(0)} &= \{ v \in V \mid x_v = 0 \} = \{ v \in V \mid z_v 
            = 0 \} = \gamma_z^{(0)}\,,  \\
            \Gamma_x^{(0)} &= \{ v \in V \mid \exists c \in \Neighbours{v} 
            \text{, s.t. } y_c = a_{cv} x_v \}   \\
                    &= \{ v \in V \mid \exists c \in \Neighbours{v} \text{, 
                    s.t. } 
                    s_c = z_v \} = \Gamma_z^{(0)} \,.
        \end{align*}
        
        {\flushleft \emph{Inductive Step}.}
        
        Consider iteration $\ell \geq 1$. First note that all $v \in V$ with $x_v = 0$ (and hence $z_v = 0$) belong to both $\gamma_x^{(\ell)}$ and $\gamma_z^{(\ell)}$.
        
        If $x_v > 0$ (and hence $z_v = 1$) then from \cref{alg:ipa:muvc} of \cref{alg:ipa} and the definition of $\gamma_x^{(\ell)}$, we have $v \in \gamma_x^{(\ell)}$ if and only if there exists $c \in \Neighbours{v}$ such that $\mu_{c \to v}^{(\ell)} = x_v$. More precisely:
        \begin{align*}
        a_{cv} x_v &= y_c - \sum_{\substack{v' \in \Neighbours[]{c} \\ v' \neq v}} a_{cv'} M_{v' \to c}^{(\ell-1)} \\
        &=a_{cv} x_v + \sum_{\substack{v' \in \Neighbours[]{c} \\ v' \neq v}} a_{cv'} \left( x_{v'} - M_{v' \to c}^{(\ell-1)} \right) \leq a_{cv} x_v \,.
        \end{align*}
        Equality holds if and only if $M_{v' \to c}^{(\ell-1)} = x_{v'}$ for all $v' \in \Neighbours[]{c} \setminus \{v\}$ or, in our notation, $\Neighbours[]{c} \setminus \{v\} \subset \Gamma_x^{(\ell-1)}$. However, by inductive assumption $\Gamma_z^{(\ell-1)} = \Gamma_x^{(\ell-1)}$ and hence $\Lambda_{v' \to c}^{(\ell-1)} = z_{v'}$ for all $v' \in 
            \Neighbours{c} \setminus \{ v \}$. This is equivalent to
            $\lambda_{c \to v}^{(\ell)} = z_v$ and thus $v \in \gamma_z^{(\ell)}$.
        
        Hence, for all $v \in V$, $v$ either belongs to both $\gamma_x^{(\ell)}$ and 
        $\gamma_z^{(\ell)}$, or to none of them.
        
        Analogously, we can show that $\Gamma_x^{(\ell)} = \Gamma_z^{(\ell)}$. Details are omitted for brevity.
    \end{IEEEproof}
    
    \cref{thm:0-1-x} gives a powerful tool for analysis of IPA performance. Instead of considering $A \in \Reals_{\geq 0}^{m \times n}$ and $\vec x \in \Reals^n_{\geq 0}$ we need only to work with binary $A$ and $\vec x$ (although all operations are still performed over $\Reals$). Thus, in the rest of the paper, we assume that $A$ is binary.

\subsection{Termatiko Sets} \label{sec:termatiko_sets}
We define termatiko sets through failures of the IPA.

\begin{define}
    We call $T \subset V$ a \emph{termatiko set} if and only if $\IPA(A \vec{x}_T, A) = \vec 0$, where $\vec x_T$ is a binary vector with support $\supp(\vec x_T) = T$.
\end{define}
From \cref{thm:0-1-x}, it follows that the IPA completely fails to recover $\vec x \in \Reals^n_{\geq 0}$ if and only if $\supp(\vec x) = T$, where $T$ is a nonempty termatiko set.

\begin{thm} \label{thm:termatikos}
    Let $T$ be a subset of the set of variable nodes $V$. We denote by $N = \Neighbours T$ the set of measurement nodes connected to $T$ and also denote by $S$ the other variable nodes connected only to $N$ as follows:
    \begin{equation} \notag %
    S = \left\{ v \in V \setminus T \,:\, \Neighbours[N] v = \Neighbours v \right\} \,.
    \end{equation}
    Then, $T$ is a termatiko set if and only if for each $c \in N$ one of the following two conditions holds (cf.\ \cref{fig:term-class1-ex,fig:term-class2-ex}):
    \begin{itemize}
        \item $c$ is connected to $S$ (this implies $S \neq \varnothing$);
        \item $c$ is not connected to $S$ and
        \begin{gather*}
            \Big|\left\{
                v \in \Neighbours[T]{c} \,:\, \forall c' \in \Neighbours v ,\, |\Neighbours[T]{c'}| \geq 2
            \right\}\Big| \geq 2 \,.
        \end{gather*}
    \end{itemize}
\end{thm}

\begin{figure}
    \centering
    \begin{tikzpicture}
    \node at (1.0,+1) [circle,draw] (v0) {$v_{0}$};
    \node at (3.0,+1) [circle,draw] (v1) {$v_{1}$};
    \node at (2.0,-1) [circle,draw] (v2) {$v_{2}$};
    \node at (6.0,-1) [circle,draw,dotted,gray] (v3) {$v_{3}$};
    \node at (4.0,-1) [circle,draw,dotted,gray] (v4) {$v_{4}$};
    \node at (5.0,-1) [circle,draw,dotted,gray] (v5) {$v_{5}$};
    \node at (3.0,-1) [circle,draw,dotted,gray] (v6) {$v_{6}$};
    
    \node at (1.0,0) [square,draw] (c0) {$c_{0}$};
    \node at (3.0,0) [square,draw] (c1) {$c_{1}$};
    \node at (5.0,0) [square,draw,dotted,gray] (c2) {$c_{2}$};
    
    \node at (0,+1) {$T:$};
    \node at (0,0) {$N:$};
    \node at (0,-1) {$S:$};
    
    \draw (c0) -- (v0);
    \draw (c0) -- (v2);
    \draw[dotted,gray] (c0) -- (v4);
    \draw[dotted,gray] (c0) -- (v6);
    
    \draw (c1) -- (v1);
    \draw (c1) -- (v2);
    \draw[dotted,gray] (c1) -- (v5);
    \draw[dotted,gray] (c1) -- (v6);
    
    \draw[dotted,gray] (c2) -- (v3);
    \draw[dotted,gray] (c2) -- (v4);
    \draw[dotted,gray] (c2) -- (v5);
    \draw[dotted,gray] (c2) -- (v6);
    \end{tikzpicture}
        \vspace{-1ex}
    \caption{Example of a termatiko set $T$ with all measurement nodes in $N$ connected to both $T$ and $S$ (cf.\ \cref{thm:termatikos}). The rest of the Tanner graph is drawn dotted.}
        \vspace{-1ex}
    \label{fig:term-class1-ex}
\end{figure}
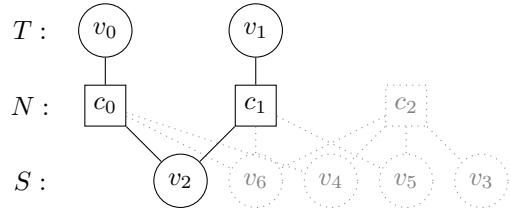

\begin{figure}
    \centering
    \begin{tikzpicture}
    \node at (1.0,-1) [circle,draw] (v0) {$v_{0}$};
    \node at (3.25,+1) [circle,draw] (v5) {$v_{5}$};
    \node at (2.0,+1) [circle,draw,thick] (v1) {$v_{1}$};
    \node at (2.5,-1) [circle,draw] (v3) {$v_{3}$};
    \node at (4.0,-1) [circle,draw] (v4) {$v_{4}$};
    \node at (4.5,+1) [circle,draw] (v2) {$v_{2}$};
    \node at (5.5,-1) [circle,draw] (v6) {$v_{6}$};
    
    \node at (1.0,0) [square,draw,thick] (c0) {$c_{0}$};
    \node at (3.25,0) [square,draw,thick] (c1) {$c_{1}$};
    \node at (5.5,0) [square,draw] (c2) {$c_{2}$};
    
    \draw (c0) -- (v0);
    \draw[ultra thick] (c0) -- (v1);
    \draw (c0) -- (v4);
    \draw (c0) -- (v6);
    
    \node at (0,+1) {$T:$};
    \node at (0,0) {$N:$};
    \node at (0,-1) {$S:$};
    
    \draw (c1) -- (v5);
    \draw[ultra thick] (c1) -- (v1);
    \draw (c1) -- (v2);
    
    \draw (c2) -- (v2);
    \draw (c2) -- (v3);
    \draw (c2) -- (v4);
    \draw (c2) -- (v5);
    \draw (c2) -- (v6);
    \end{tikzpicture}
        \vspace{-1ex}
    \caption{Example of a termatiko set $T$ with a measurement node $c_1$ connected to $T$ only (cf.\ \cref{thm:termatikos}). Highlighted is the connection to a measurement node $c_0$, which is connected to $T$ only once.}
        \vspace{-1ex}
    \label{fig:term-class2-ex}
\end{figure}
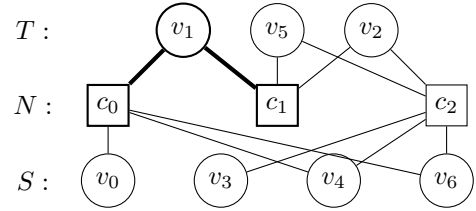

\begin{IEEEproof}
    Consider the problem $\IPA(A \vec{x}_T, A)$, where $\vec x_T$ is a binary vector with support $\supp(\vec x_T) = T$.
    
    We first note that measurement nodes in $C \setminus N$ have value zero and hence all variable nodes connected to them (i.e., $v \in V \setminus (T \cup S)$) are recovered with zeros at the initialization step of \cref{alg:ipa}. As a consequence, they can be safely pruned and w.l.o.g.\ we can assume that $C = N$ and $V = T \cup S$.
    
    Assume that $T$ satisfies the conditions of the theorem and consider the problem  $\IPA(A \vec{x}_T, A)$. %
We show by induction that for all $v \in T \cup S$ at each iteration $\ell \geq 0$ it holds that $\mu_{v \to \cdot}^{(\ell)} = 0$ and $M_{v \to \cdot}^{(\ell)} \geq 1$. Moreover, each measurement node $c \in N$ that is not connected to $S$ has at least two different neighbors $v_1, v_2 \in T$ with $M_{v_1 \to \cdot}^{(\ell)} \geq 2$ and $M_{v_2 \to \cdot}^{(\ell)} \geq 2$.
    
    {\flushleft \emph{Base Case.}}
    
    For $\ell = 0$ we immediately obtain from \cref{alg:ipa} that $\mu_{v \to \cdot}^{(0)} = 0$ and, as each $c \in N$ has at least one nonzero neighbor, $M_{v \to \cdot}^{(0)} \geq 1$. In addition, consider $c \in N$ that is not connected to $S$. It has at least two different neighbors $v_1, v_2 \in T$, each connected only to measurement nodes with not less than two neighbors in $T$. Therefore, $M_{v_1 \to \cdot} \geq 2$ and $M_{v_2 \to \cdot} \geq 2$. %
    
    {\flushleft \emph{Inductive Step.}}
    
    Consider $\ell \geq 1$. For all $c \in N$ and all $v \in \Neighbours c$,
    \[
        M_{c \to v}^{(\ell)} = y_c - \sum_{v' \in \Neighbours{c} \,, v' \neq v} \mu_{v' \to c}^{(\ell-1)} = y_c \,.
    \]
    Hence, upper bounds are exactly the same as for $l = 0$ and the same inequalities hold for them.
    
    In order to find lower bounds, we consider two cases for $c \in N$. If $c$ is connected to $S$, then
        \begin{multline*}
            y_c - \sum_{\substack{v' \in \Neighbours{c} \\ v' \neq v}} M_{v' \to c}^{(\ell-1)} 
            \leq \left( |\Neighbours{c}| - 1 \right) - \sum_{\substack{v' \in \Neighbours{c} \\ v' \neq v}} 1 = 0
        \end{multline*}
    and therefore $\mu_{c \to v}^{(\ell)} = 0$. If $c$ is connected to $T$ only, then
        \begin{multline*}
            y_c - \sum_{\substack{v' \in \Neighbours{c} \\ v' \neq v}} M_{v' \to c}^{(\ell-1)} 
            \leq |\Neighbours[T]{c}| - \Bigg( 1 + \sum_{\substack{v' \in \Neighbours[T]{c} \\ v' \neq v}} 1 \Bigg) = 0
        \end{multline*}
    and again $\mu_{c \to v}^{(\ell)} = 0$. Here, the extra $1$ inside the parenthesis indicates the fact that for at least one $v'$ we have $M_{v' \to c}^{(\ell-1)} \geq 2$. Thus, at each iteration of the IPA for each $v \in V$ the lower bound is equal to zero, and the algorithm will return $\hat{\vec x} = \vec 0$.
     
    We have demonstrated that if $T$ satisfies the conditions of the theorem, it is a termatiko set. What remains to be proven is that if $T$ does not satisfy the conditions of the theorem, the IPA can recover at least some of the nonzero values. 
    
    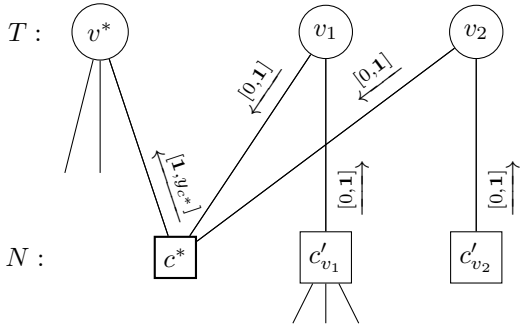
\begin{figure}
        \centering
        \begin{tikzpicture}
        \node at (-1, 3) {$T:$};
        \node at (-1, 0) {$N:$};
        
        \node at (0, 3) [circle,draw] (vstar) {$v^*$};
        \node at (3, 3) [circle,draw] (v1) {$v_1$};
        \node at (5, 3) [circle,draw] (v2) {$v_2$};

        \node at (1, 0) [square,draw,thick] (cstar) {$c^*$};
        \node at (3, 0) [square,draw] (c1) {$c'_{v_1}$};
        \node at (5, 0) [square,draw] (c2) {$c'_{v_2}$};
        
        \draw (cstar) -- (vstar);
        \draw (cstar) -- (vstar) node [near start,above,sloped] (mcsvs) {$\xleftarrow{[\bm 1,y_{c^*}]}$};
        \draw (v1) -- (cstar);
        \draw (v1) -- (cstar) node [near start,above,sloped] (mv1cs) {$\xleftarrow{[0,\bm 1]}$};
        \draw (v2) -- (cstar);
        \draw (v2) -- (cstar) node [near start,above,sloped] (mv2cs) {$\xleftarrow{[0,\bm 1]}$};
        
        \draw (c1) -- (v1);
        \draw (c1) -- (v1) node [near start,below,sloped] (mc1v1) {$\xrightarrow{[0,\bm 1]}$};

        \draw (c2) -- (v2);
        \draw (c2) -- (v2) node [near start,below,sloped] (mc2v2) {$\xrightarrow{[0,\bm 1]}$};
        
        \node at (-.5, 1) (f1) {};
        \node at (0, 1)   (f2) {};
        \draw (vstar) -- (f1);
        \draw (vstar) -- (f2);
        
        \node at (2.5, -1) (f3) {};
        \node at (3, -1) (f4) {};
        \node at (3.5, -1) (f5) {};
        \draw (c1) -- (f3);
        \draw (c1) -- (f4);
        \draw (c1) -- (f5);
        \end{tikzpicture}
        \vspace{-2ex}
        \caption{Exact bounds propagation in a nontermatiko set. Here $[\mu,M]$ denotes sending a lower bound of $\mu$ and an upper bound of $M$ in the direction given by the corresponding arrow. Numbers in bold are exact bounds.}
        \vspace{-2ex}
        \label{fig:non-termatiko}
    \end{figure}
    
    Assume that there exists $c^* \in N$ not connected to $S$ (i.e., $\Neighbours[T]{c^*} = \Neighbours[]{c^*}$) and such that 
    \[
        \Big|\left\{
            v \in \Neighbours[T]{c^*} \,:\, \forall c' \in \Neighbours v ,\, |\Neighbours[T]{c'}| \geq 2
        \right\}\Big| \leq 1 \,.
    \]
    If this set has one element, denote it by $v^*$. If it is empty, let $v^*$ be any element of $\Neighbours[T]{c^*}$.
    
    A special case when $|\Neighbours[T]{c^*}| = 1$ is trivial. Otherwise, for any $v \in \Neighbours[T]{c^*} \setminus \{ v^* \}$, there exists $c'_v \in \Neighbours{v}$ such that $|\Neighbours[T]{c'_v}| \leq 1$, which in truth means that $\Neighbours[T]{c'_v} = \{ v \}$.
    
    Hence, at the initialization step of the IPA, for all $v \in \Neighbours[T]{c^*} \setminus \{ v^* \}$ we will have $\mu_{v \to \cdot}^{(0)} = 0$ and $M_{v \to \cdot}^{(0)} = 1$. Therefore, at iteration $\ell = 1$:
    \[
    \mu^{(1)}_{c^* \to v^*} \leftarrow y_{c^*} - \sum_{\substack{v \in \Neighbours[T]{c^*} \\ v \neq v^*}} M^{(0)}_{v \to c^*} = y_{c^*} - \sum_{\substack{v \in \Neighbours[T]{c^*} \\ v \neq v^*}} 1 = 1 \,.
    \]
    Thus, the IPA will output 1 for position $v^* \in T$, which means that $T$ is not a termatiko set. See \cref{fig:non-termatiko} for illustration.
\end{IEEEproof}

\cref{thm:termatikos} gives a precise graph-theoretic description of termatiko sets. In fact, it defines two important subclasses of termatiko sets; stopping sets and sets with all $c \in N$ connected to both $T$ and $S$. Also, it is worth noting that $T \cup S$ is a stopping set. Thus, a termatiko set is always a subset of some stopping set. We define the size of the smallest nonempty termatiko set as the \emph{termatiko distance}.
    \begin{figure}
        \centering
        \begin{tikzpicture}
        \node at (0.0,+1.5) [circle,draw,thick,fill=black!10] (v1) {$v_{1}$};
        \node at (0.0,2.1) {0};
        \node at (1.5,+1.5) [circle,draw,thick,fill=black!10] (v2) {$v_{2}$};
        \node at (1.5,2.1) {0};
        \node at (3.0,+1.5) [circle,draw,thick,fill=black!10] (v3) {$v_{3}$};
        \node at (3.0,2.1) {1};
        \node at (4.5,+1.5) [circle,draw,thick,fill=black!10] (v4) {$v_{4}$};
        \node at (4.5,2.1) {1};
        \node at (6.0,+1.5) [circle,draw] (v5) {$v_{5}$};
        \node at (6.0,2.1) {0};
        \node at (7.5,+1.5) [circle,draw] (v6) {$v_{6}$};
        \node at (7.5,2.1) {0};
        
        \node at (0.75,0) [rectangle,draw] (c1) {$c_{1}$};
        \node at (0.75,-.5) {0};
        \node at (2.75,0) [rectangle,draw] (c2) {$c_{2}$};
        \node at (2.75,-.5) {1};
        \node at (4.75,0) [rectangle,draw] (c3) {$c_{3}$};
        \node at (4.75,-.5) {1};
        \node at (6.75,0) [rectangle,draw] (c4) {$c_{4}$};
        \node at (6.75,-.5) {2};
        
        \draw (c1) -- (v1);
        \draw (c1) -- (v2);
        \draw (c1) -- (v6);
        
        \draw (c2) -- (v1);
        \draw (c2) -- (v4);
        \draw (c2) -- (v5);
        
        \draw (c3) -- (v2);
        \draw (c3) -- (v3);
        \draw (c3) -- (v6);
        
        \draw (c4) -- (v3);
        \draw (c4) -- (v4);
        \draw (c4) -- (v5);
        
        \end{tikzpicture}
        \vspace{-2ex}
        \caption{Counter-example to \cite[Thm.~2]{RDVD12}. The set of variable nodes is $V = \{v_1,\dotsc,v_6\}$ (circles) and the set of measurement nodes is $C = \{c_1,\dotsc,c_4\}$ (squares).  The integer attached to a node is its corresponding value ($x_{v_i}$ for variable node $v_i$ and $y_{c_i}$ for measurement node $c_i$). $V_S = \{ v_1,v_2,v_3,v_4\} \subset V$  (shaded in gray) is a minimal stopping set and $c_1$ is a zero-valued ($y_{c_1}=0$) measurement node connected to $V_S$.  Note that $v_5$ is not in $V_S$, but exactly because of it, the IPA cannot correctly recover $v_4$.}
        \vspace{-2ex}
        \label{fig:counter-th2} 
    \end{figure}
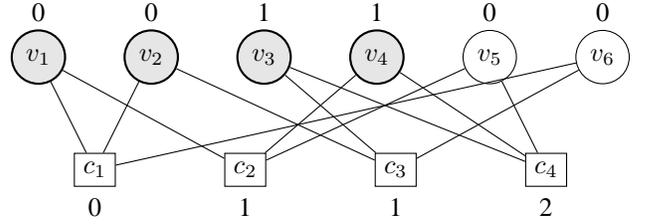

\subsection{General Failing Sets} \label{sec:criterion}

In \cref{sec:termatiko_sets}, we defined termatiko sets as supports of binary vectors that avert the IPA from recovering any of the ones. However, the algorithm can recover only some of the positions of ones. The next proposition (proof omitted) gives a connection between (partial) failures of the IPA and termatiko sets; it shows that the IPA fails on any signal in $\Reals^n_{\geq 0}$ if and only if its support contains a nonempty termatiko set.

\begin{table*}[!t]
    \scriptsize \centering \caption{Estimated termatiko set size spectra (initial part) of measurement matrices from \cref{sec:numerical_results}, where $\hat{h}_{\rm min}$ denotes the estimated termatiko distance. $\mathfrak T_1$ corresponds to termatiko sets with all measurement nodes in $N$ connected to both $T$ and $S$, and $\mathfrak T_2$ corresponds to all the remaining termatiko sets. Also shown are the exact stopping distances and stopping set size spectra (initial part). Entries in bold are exact values. For  $A^{(1)}$, the heuristic approach gives a multiplicity of $5875518$ for size $5$, while the exact number is $6318378$ (an underestimation of about $7.5\%$).} \label{table_of_codes}
    \def\Hline{\noalign{\hrule height 2\arrayrulewidth}}
    \vskip -3.0ex %
    \begin{tabular}{ccccc}
        \Hline \\ [-2.0ex]
        Measurement matrix & $\hat{h}_{\min}$ & Initial estimated termatiko set size spectrum & $s_{\min}$ & Initial stopping set size spectrum \\
        \hline
        \\ [-2.0ex] \hline  \\ [-2.0ex]
        $A^{(1)}$  & $\bm 3$ & $\mathfrak T_1$: $(\bm{3630},\bm{93775},\bm{6318378},48548225,71709440,$ \\
        & & $36514170,7969060,856801,41745)$ & $\bm 6$ & $(\bm{1815},\bm{605},\bm{45375},\bm{131890},\bm{3550382},\bm{28471905})$\\
        & & $\mathfrak T_2$: $(0,0,0,410190,18610405,71153445,86844725,$\\
        & & $58849681,28430160)$ \\
        \hline \\[-2.0ex]
        $A^{(2)}$ & $9$ & $\mathfrak T_1$: $(465, 3906, 12555, 8835, 0, 0, \dotsc)$  & $\bm{18}$ & $(\bm{465},\bm{2015},\bm{9548},\bm{23715},\bm{106175})$ \\
        & & $\mathfrak T_2$: $(0,0,0,1860, 5115, 10695, 2325, 5580, 2325, 6045$ \\
        & & $10850, 22103, 39990, 106175)$ \\
        \hline \\[-2.0ex]
        $A^{(3)}$  & $8$ & $\mathfrak T_1$: $(228, 0, 0, \dotsc)$ & $\bm 9$ & $(\bm{76},\bm 0,\bm 0,\bm 0,\bm{76},\bm{76},\bm{304},\bm{1520})$\\
        & & $\mathfrak T_2$: $(0, 76, 0, 76, 684, 532, 152, 532, 1520)$ \\
        \hline \\[-2.0ex]
        $A^{(4)}$  & $8$ & $\mathfrak T_1$: $(184, 598, 1242, 391, 0, 0)$ & $\bm{15}$ & $(\bm{46},\bm{161},\bm{391},\bm{897},\bm{2093},\bm{5796})$\\
        & & $\mathfrak T_2$: $(0, 0, 0, 69, 23, 0, 23, 46, 161, 391, 1012, 2300, 5796)$ \\
        \hline \\[-2.0ex]
        $A^{(5)}$  & $7$ & $\mathfrak T_1$: $(106, 0, 0, 53, 901, 3233, 954, 53, 0, 0, \dotsc)$ & $\bm{14}$ & $(\bm{53},\bm 0,\bm 0,\bm 0,\bm 0,\bm{53},\bm{106},\bm{583},\bm{1484},\bm{3922},\bm{9964})$\\
        & & $\mathfrak T_2$: $(0, 0, 0, 0, 0, 0, 106, 265, 106, 636, 689, 477, $ \\
        & & $583, 371, 1325, 2915, 5830, 9964)$ \\
        \hline
    \end{tabular}
    \vskip -3.5ex %
\end{table*}

\begin{proposition}
    The IPA fails to fully recover a nonnegative real signal $\vec x \in \Reals^n_{\geq 0}$ if and only if the support of $\vec x$ contains a nonempty termatiko set.
\end{proposition}

%
%
%

%

\section{Counter-Example to \cite[Thm.~2]{RDVD12}}

In \cite[Thm.~2]{RDVD12}, a condition for full recovery of $\vec x$ is stated. However, in \cref{fig:counter-th2}, we provide a counter-example to this theorem. Note that the Tanner graph of \cref{fig:counter-th2} is $(2,3)$-regular (only regular Tanner graphs with variable node degree at least two were considered in \cite{RDVD12}) and satisfies the conditions of \cite[Thm.~2]{RDVD12}. In particular, there are at most $|V_S|-2=2$ nonzero-valued variable nodes which are both in $V_S$ ($V_S$ is a minimal stopping set contained in $V$); and there is at least one zero-valued measurement node among the neighbors of $V_S$. However, it can be readily seen that the IPA will output $\hat{\vec x} = (0,0,1,0,0,0)$, i.e., it recovers only one nonzero variable node ($v_4$ and $v_5$ are both connected to $c_2$ and $c_4$ and thus indistinguishable; hence, the IPA will definitely fail).
    We believe that the main problematic issue in the proof given in \cite{RDVD12} is that variable nodes 
    outside of the minimal stopping set $V_S$ are not considered. Despite the fact that such
    nodes will be recovered as zeros in the end (because of the specific implementation of the IPA, see \cref{alg:hatxv}), during iterations they still can ``disturb'' the values inside of the stopping set.

Finally, we remark that since the statement of \cite[Thm.~2]{RDVD12} is used in the  proof of \cite[Thm.~3]{RDVD12}, it should be further verified.

%
%
%
%
%
%
%
%

%
%

%
%

%
%

%
%

%
%
%

\section{Heuristic to Find Small-Size Termatiko Sets} \label{sec:alg_small_size}

As shown in \cref{sec:failing_sets}, (small-size) stopping sets may contain termatiko sets as proper subsets. Thus, one way to locate termatiko sets is to first enumerate all stopping sets of size at most $\tau$ (for a given binary measurement matrix and threshold $\tau$) and then look for subsets that are termatiko sets. %
%
%
For a given binary measurement matrix $A$, small-size stopping sets can be identified using the algorithm from \cite{ros12,ros09}.

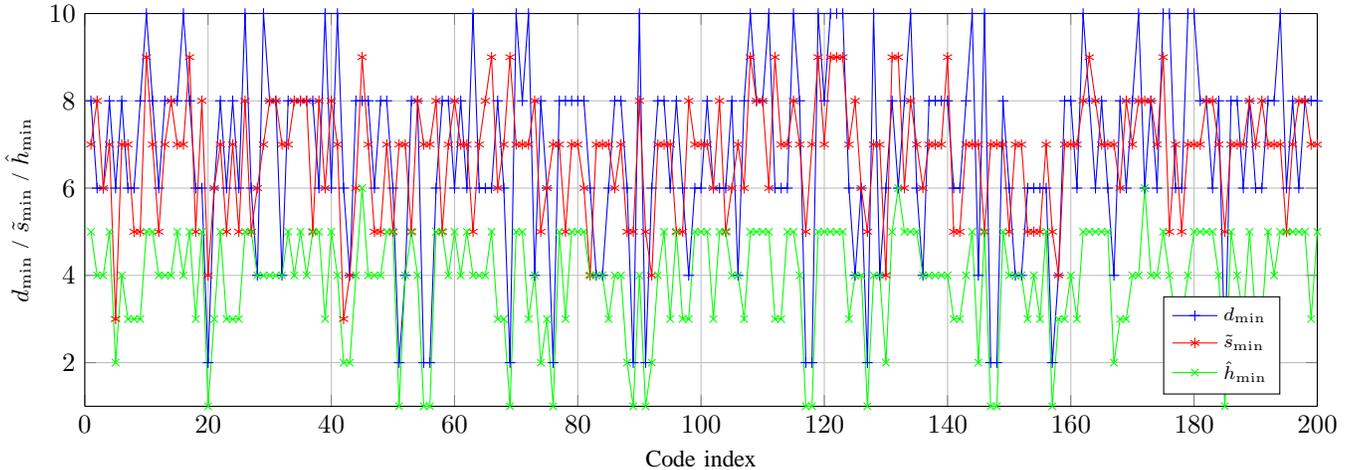
\begin{figure*}[tbp]
\centering
%
%
\begin{tikzpicture}[font=\small]

\begin{axis}[%
width=1.85\columnwidth,
height=0.59\columnwidth,
at={(0.808889in,0.513333in)},
scale only axis,
separate axis lines,
every outer x axis line/.append style={black},
every x tick label/.append style={font=\color{black}},
xmin=0,
xmax=200,
xlabel={Code index},
xmajorgrids,
every outer y axis line/.append style={black},
every y tick label/.append style={font=\color{black}},
ymin=1,
ymax=10,
ylabel={$d_{\rm min}$ / $\tilde{s}_{\rm min}$ / $\hat{h}_{\rm min}$},
ymajorgrids,
legend style={at={(0.97,0.03)},anchor=south east,legend cell align=left,align=left,draw=black,font=\scriptsize}
]
\addplot [color=blue,solid,mark=+,mark options={solid}]
  table[row sep=crcr]{%
1	8\\
2	6\\
3	6\\
4	8\\
5	6\\
6	8\\
7	6\\
8	6\\
9	8\\
10	10\\
11	8\\
12	6\\
13	8\\
14	8\\
15	8\\
16	10\\
17	8\\
18	6\\
19	6\\
20	2\\
21	6\\
22	8\\
23	6\\
24	8\\
25	6\\
26	10\\
27	6\\
28	4\\
29	10\\
30	8\\
31	8\\
32	4\\
33	8\\
34	8\\
35	8\\
36	8\\
37	8\\
38	6\\
39	10\\
40	6\\
41	10\\
42	6\\
43	4\\
44	8\\
45	8\\
46	8\\
47	6\\
48	8\\
49	8\\
50	6\\
51	2\\
52	4\\
53	8\\
54	8\\
55	2\\
56	2\\
57	6\\
58	8\\
59	8\\
60	6\\
61	8\\
62	6\\
63	10\\
64	6\\
65	6\\
66	6\\
67	8\\
68	6\\
69	2\\
70	10\\
71	8\\
72	10\\
73	4\\
74	8\\
75	6\\
76	2\\
77	8\\
78	8\\
79	8\\
80	8\\
81	8\\
82	6\\
83	4\\
84	4\\
85	6\\
86	8\\
87	8\\
88	6\\
89	2\\
90	10\\
91	2\\
92	6\\
93	8\\
94	8\\
95	6\\
96	8\\
97	6\\
98	4\\
99	6\\
100	6\\
101	8\\
102	6\\
103	6\\
104	6\\
105	8\\
106	4\\
107	8\\
108	10\\
109	8\\
110	8\\
111	10\\
112	6\\
113	6\\
114	6\\
115	10\\
116	8\\
117	2\\
118	2\\
119	10\\
120	8\\
121	10\\
122	10\\
123	10\\
124	6\\
125	4\\
126	6\\
127	2\\
128	10\\
129	4\\
130	6\\
131	8\\
132	6\\
133	8\\
134	10\\
135	6\\
136	4\\
137	8\\
138	8\\
139	8\\
140	8\\
141	6\\
142	6\\
143	8\\
144	10\\
145	4\\
146	10\\
147	2\\
148	2\\
149	8\\
150	6\\
151	4\\
152	4\\
153	6\\
154	6\\
155	6\\
156	6\\
157	2\\
158	4\\
159	8\\
160	8\\
161	6\\
162	10\\
163	8\\
164	6\\
165	8\\
166	6\\
167	4\\
168	8\\
169	6\\
170	8\\
171	10\\
172	6\\
173	8\\
174	6\\
175	10\\
176	10\\
177	6\\
178	6\\
179	10\\
180	10\\
181	8\\
182	8\\
183	6\\
184	8\\
185	2\\
186	8\\
187	8\\
188	6\\
189	8\\
190	6\\
191	6\\
192	8\\
193	8\\
194	10\\
195	6\\
196	8\\
197	6\\
198	8\\
199	8\\
200	8\\
};
\addlegendentry{$d_{\rm min}$};

\addplot [color=red,solid,mark=asterisk,mark options={solid}]
  table[row sep=crcr]{%
1	7\\
2	8\\
3	6\\
4	7\\
5	3\\
6	7\\
7	7\\
8	5\\
9	5\\
10	9\\
11	7\\
12	5\\
13	7\\
14	8\\
15	7\\
16	7\\
17	9\\
18	5\\
19	8\\
20	4\\
21	6\\
22	7\\
23	5\\
24	7\\
25	5\\
26	8\\
27	5\\
28	6\\
29	7\\
30	8\\
31	8\\
32	7\\
33	7\\
34	8\\
35	8\\
36	8\\
37	5\\
38	8\\
39	6\\
40	8\\
41	7\\
42	3\\
43	4\\
44	6\\
45	9\\
46	7\\
47	5\\
48	5\\
49	7\\
50	5\\
51	7\\
52	7\\
53	5\\
54	8\\
55	7\\
56	7\\
57	8\\
58	5\\
59	7\\
60	8\\
61	7\\
62	7\\
63	5\\
64	7\\
65	8\\
66	9\\
67	6\\
68	7\\
69	9\\
70	7\\
71	7\\
72	7\\
73	8\\
74	5\\
75	6\\
76	7\\
77	7\\
78	5\\
79	7\\
80	7\\
81	6\\
82	4\\
83	7\\
84	7\\
85	7\\
86	6\\
87	7\\
88	5\\
89	5\\
90	8\\
91	5\\
92	4\\
93	7\\
94	7\\
95	7\\
96	5\\
97	5\\
98	8\\
99	7\\
100	7\\
101	7\\
102	6\\
103	8\\
104	5\\
105	6\\
106	7\\
107	6\\
108	9\\
109	8\\
110	8\\
111	6\\
112	9\\
113	7\\
114	7\\
115	8\\
116	7\\
117	5\\
118	7\\
119	9\\
120	7\\
121	9\\
122	9\\
123	9\\
124	7\\
125	8\\
126	6\\
127	5\\
128	7\\
129	7\\
130	4\\
131	9\\
132	9\\
133	6\\
134	8\\
135	7\\
136	6\\
137	7\\
138	7\\
139	7\\
140	9\\
141	5\\
142	5\\
143	7\\
144	7\\
145	7\\
146	5\\
147	7\\
148	7\\
149	7\\
150	5\\
151	7\\
152	7\\
153	5\\
154	5\\
155	5\\
156	7\\
157	5\\
158	4\\
159	7\\
160	7\\
161	7\\
162	8\\
163	9\\
164	8\\
165	7\\
166	7\\
167	7\\
168	6\\
169	8\\
170	7\\
171	8\\
172	8\\
173	8\\
174	7\\
175	9\\
176	5\\
177	7\\
178	5\\
179	7\\
180	7\\
181	7\\
182	8\\
183	8\\
184	7\\
185	5\\
186	7\\
187	7\\
188	7\\
189	8\\
190	7\\
191	8\\
192	7\\
193	7\\
194	7\\
195	5\\
196	7\\
197	8\\
198	8\\
199	7\\
200	7\\
};
\addlegendentry{$\tilde{s}_{\rm min}$};

\addplot [color=green,solid,mark=x,mark options={solid}]
  table[row sep=crcr]{%
1	5\\
2	4\\
3	4\\
4	5\\
5	2\\
6	4\\
7	3\\
8	3\\
9	3\\
10	5\\
11	5\\
12	4\\
13	4\\
14	4\\
15	5\\
16	4\\
17	5\\
18	3\\
19	5\\
20	1\\
21	3\\
22	5\\
23	3\\
24	3\\
25	3\\
26	5\\
27	5\\
28	4\\
29	4\\
30	4\\
31	4\\
32	4\\
33	5\\
34	4\\
35	5\\
36	4\\
37	5\\
38	5\\
39	3\\
40	5\\
41	4\\
42	2\\
43	2\\
44	4\\
45	6\\
46	4\\
47	4\\
48	4\\
49	5\\
50	5\\
51	1\\
52	4\\
53	5\\
54	4\\
55	1\\
56	1\\
57	5\\
58	5\\
59	4\\
60	5\\
61	4\\
62	5\\
63	4\\
64	4\\
65	4\\
66	5\\
67	3\\
68	3\\
69	1\\
70	5\\
71	5\\
72	3\\
73	4\\
74	2\\
75	3\\
76	1\\
77	5\\
78	3\\
79	5\\
80	5\\
81	5\\
82	4\\
83	4\\
84	4\\
85	3\\
86	4\\
87	4\\
88	2\\
89	1\\
90	4\\
91	1\\
92	2\\
93	4\\
94	5\\
95	3\\
96	5\\
97	3\\
98	3\\
99	5\\
100	5\\
101	5\\
102	3\\
103	5\\
104	5\\
105	4\\
106	4\\
107	3\\
108	5\\
109	5\\
110	5\\
111	5\\
112	3\\
113	3\\
114	5\\
115	5\\
116	4\\
117	1\\
118	1\\
119	5\\
120	5\\
121	5\\
122	5\\
123	5\\
124	3\\
125	4\\
126	4\\
127	1\\
128	4\\
129	4\\
130	2\\
131	5\\
132	6\\
133	5\\
134	5\\
135	5\\
136	4\\
137	4\\
138	4\\
139	4\\
140	4\\
141	3\\
142	3\\
143	4\\
144	5\\
145	2\\
146	5\\
147	1\\
148	1\\
149	5\\
150	4\\
151	4\\
152	4\\
153	3\\
154	4\\
155	3\\
156	5\\
157	1\\
158	3\\
159	3\\
160	4\\
161	3\\
162	5\\
163	5\\
164	5\\
165	5\\
166	5\\
167	2\\
168	3\\
169	3\\
170	4\\
171	4\\
172	6\\
173	4\\
174	4\\
175	5\\
176	4\\
177	3\\
178	3\\
179	4\\
180	5\\
181	5\\
182	5\\
183	5\\
184	4\\
185	1\\
186	5\\
187	4\\
188	3\\
189	5\\
190	3\\
191	3\\
192	5\\
193	4\\
194	5\\
195	5\\
196	5\\
197	5\\
198	5\\
199	3\\
200	5\\
};
\addlegendentry{$\hat{h}_{\rm min}$};

\end{axis}
\end{tikzpicture}%
\vspace{-2ex}
\caption{Minimum distance $d_{\rm min}$, minimum size of a noncodeword stopping set $\tilde{s}_{\rm min}$, and estimated termatiko distance $\hat{h}_{\rm min}$ versus code index for randomly generated binary measurement matrices from a protograph-based $(3,6)$-regular LDPC code ensemble.}
\label{fig:33_ldpc_ensemble}
    \vskip -0.75ex %
\end{figure*}

\section{Numerical Results} \label{sec:numerical_results}

In this section, we present numerical results for different specific measurement matrices and also for ensembles of measurement matrices. For all matrices we first find all stopping sets of size less than some threshold using the algorithm from  \cite{ros12,ros09}. Then, we exhaustively search for termatiko sets as subsets of these stopping sets as explained in \cref{sec:alg_small_size}. The results are tabulated in \cref{table_of_codes} for five different measurement matrices, denoted by $A^{(1)}$, $A^{(2)}$, $A^{(3)}$, $A^{(4)}$, and $A^{(5)}$, respectively.  Due to the heuristic nature of the approach, the estimated termatiko distance is a true upper bound on the actual termatiko distance, while the estimated multiplicities are true lower bounds on the actual multiplicities. %
Measurement matrix $A^{(1)}$ is a $33 \times 121$ parity-check matrix of an array-based LDPC code of column weight $3$ and row weight $11$ \cite{fan00}, $A^{(2)}$ is the parity-check matrix of the $(155,64)$ Tanner code from \cite{tan01}, $A^{(3)}$ is taken from the IEEE802.16e standard %
(it is the parity-check matrix of a rate-$3/4$, length-$1824$ LDPC code; using base model matrix A and the alternative construction, see \cite[Eq.~(1)]{ros12}),  $A^{(4)}$ is a $276 \times 552$ parity-check matrix of an irregular LDPC code, while $A^{(5)}$ is a $159 \times 265$ parity-check matrix of a $(3,5)$-regular LDPC code built from arrays of permutation matrices from Latin squares. For the matrix $A^{(1)}$, we have also compared the results with an exact enumeration of all termatiko sets of size at most $5$. When considering all stopping sets of size at most $11$,  the heuristic approach finds the exact multiplicities  for sizes $3$ and $4$, but it underestimates  the number of termatiko sets of size $5$ by about $7.5\%$  (the missing ones are subsets of stopping sets of size $12$ to $14$),   %
which indicates that higher order terms (for all tabulated matrices) are mostly likely strict lower bounds on the exact multiplicities. 
As can be seen from the table, for all matrices except $A^{(3)}$, the estimated termatiko distance is about half the stopping distance.  Also, the smallest-size termatiko sets all correspond to termatiko sets with all measurement nodes in $N$ connected to both $T$ and $S$ %
(cf.\ \cref{thm:termatikos}). Note that the matrix $A^{(1)}$ is from a family of array-based column-weight $3$ matrices, parametrized  by an odd prime $p$. In the general case, the number of columns is $p^2$, while the number of rows in $3p$ \cite{fan00}. It is known that the minimum distance (the measurement matrix is regarded as the parity-check matrix of an LDPC code) for $p \geq 5$ is $6$ \cite[Thm.~3]{yan03}. Using the specific structure of the \emph{support matrix} of codewords of weight $6$ (see \cite[Thm.~4]{yan03}), it can be shown that there always exist termatiko sets of size $3$ for $p \geq 5$, and also that this is the smallest possible size. %
Thus, the family of parity-check matrices of array-based LDPC codes of column weight $3$ is an example of a family of measurement matrices in which the termatiko distance is exactly half the minimum distance. 

Now, consider the protograph-based $(3,6)$-regular LDPC code ensemble defined by the \emph{protomatrix} $\bm H = (3,3)$. We randomly generated $200$ parity-check matrices from this ensemble using a lifting factor of $100$ (the two nonzero entries in the protomatrix are replaced by random row-weight $3$ \emph{circulants} (each row is a right-shift of the row above it) of size $100 \times 100$). For each lifted matrix, we first found all stopping sets of size at most $16$ using the algorithm from \cite{ros12,ros09}. Then, the termatiko distance was estimated for each matrix as explained above. The results are depicted in \cref{fig:33_ldpc_ensemble} as a function of the code index (the blue curve shows the minimum distance $d_{\rm min}$, the red curve shows the minimum size of a noncodeword stopping set, denoted by $\tilde{s}_{\rm min}$, while the green curve shows the estimated termatiko distance $\hat{h}_{\rm min}$). %
The average $d_{\rm min}$, $s_{\rm min}$, and $\hat{h}_{\rm min}$ (over the $200$ matrices)  are $6.84$, $5.92$, and $3.90$, respectively. We repeated a similar experiment using a lifting factor of $200$ in which case the average $d_{\rm min}$, $s_{\rm min}$, and $\hat{h}_{\rm min}$ (again over $200$ randomly generated matrices) became  $9.21$, $7.75$, and $5.80$, respectively.

We remark that similar results (not included here) as the ones depicted in \cref{fig:33_ldpc_ensemble} have been obtained for other ensembles of measurement matrices as well. For instance, both a family of irregular rate-$1/2$ LDPC codes ($1000$ codes from the family have been considered) and the rate-$1/2$ accumulate repeat jagged accumulate ensemble from \cite{div05} show similar behaviour. %

%

%

%

%

%

%

%

%

%

%
%
%
%
%

%
%
%
%
%
%
%

\section{Conclusion} \label{sec:conclu}
In this work, we have introduced termatiko sets and shown that the IPA fails to fully recover a nonnegative real signal $\vec x \in \Reals_{\geq 0}^n$ if and only if the support of $\vec x$ contains a nonempty termatiko set, thus giving a complete (graph-theoretic) description of the failing sets of the IPA. An extensive numerical study was presented showing that having a termatiko distance strictly smaller than the stopping distance is not uncommon. In some cases, the termatiko distance can be as low as half the stopping distance. Thus, a measurement matrix (for the IPA)  should be designed to avoid small-size termatiko sets, which is considered as future work.

\balance

\end{document}